\errorstopmode
\input amssym.def
\input amssym.tex


\magnification=\magstephalf
\hsize=14.0 true cm
\vsize=19 true cm
\hoffset=1.0 true cm
\voffset=2.0 true cm

\abovedisplayskip=12pt plus 3pt minus 3pt
\belowdisplayskip=12pt plus 3pt minus 3pt
\parindent=1.0em


\font\sixrm=cmr6
\font\eightrm=cmr8
\font\ninerm=cmr9

\font\sixi=cmmi6
\font\eighti=cmmi8
\font\ninei=cmmi9

\font\sixsy=cmsy6
\font\eightsy=cmsy8
\font\ninesy=cmsy9

\font\sixbf=cmbx6
\font\eightbf=cmbx8
\font\ninebf=cmbx9

\font\eightit=cmti8
\font\nineit=cmti9

\font\eightsl=cmsl8
\font\ninesl=cmsl9

\font\sixss=cmss8 at 8 true pt
\font\sevenss=cmss9 at 9 true pt
\font\eightss=cmss8
\font\niness=cmss9
\font\tenss=cmss10

\font\sixmib=cmmib6
\font\sevenmib=cmmib7
\font\eightmib=cmmib8
\font\ninemib=cmmib9
\font\tenmib=cmmib10

 at 12 true pt
 at 12 true pt
\font\bigrm=cmr10 at 12 true pt
 at 12 true pt
 at 12 true pt

 at 16 true pt
 at 16 true pt
\font\Bigrm=cmr12 at 16 true pt
 at 16 true pt
 at 16 true pt

\catcode`@=11
\newfam\ssfam
\newfam\mibfam

\def\tenpoint{\def\rm{\fam0\tenrm}%
    \textfont0=\tenrm \scriptfont0=\sevenrm \scriptscriptfont0=\fiverm
    \textfont1=\teni  \scriptfont1=\seveni  \scriptscriptfont1=\fivei
    \textfont2=\tensy \scriptfont2=\sevensy \scriptscriptfont2=\fivesy
    \textfont3=\tenex \scriptfont3=\tenex   \scriptscriptfont3=\tenex
    \textfont\itfam=\tenit                  \def\it{\fam\itfam\tenit}%
    \textfont\slfam=\tensl                  \def\sl{\fam\slfam\tensl}%
    \textfont\bffam=\tenbf \scriptfont\bffam=\sevenbf
                           \scriptscriptfont\bffam=\fivebf
                           \def\bf{\fam\bffam\tenbf}%
    \textfont\ssfam=\tenss \scriptfont\ssfam=\sevenss
                           \scriptscriptfont\ssfam=\sevenss
                           \def\ss{\fam\ssfam\tenss}%
    \textfont\mibfam=\tenmib \scriptfont\mibfam=\sevenmib
                             \scriptscriptfont\mibfam=\sevenmib
                             \def\mib{\fam\mibfam\tenmib}%
    \normalbaselineskip=13pt
    \setbox\strutbox=\hbox{\vrule height8.5pt depth3.5pt width0pt}%
    \let\big=\tenbig
    \normalbaselines\rm}

\def\ninepoint{\def\rm{\fam0\ninerm}%
    \textfont0=\ninerm      \scriptfont0=\sixrm
                            \scriptscriptfont0=\fiverm
    \textfont1=\ninei       \scriptfont1=\sixi
                            \scriptscriptfont1=\fivei
    \textfont2=\ninesy      \scriptfont2=\sixsy
                            \scriptscriptfont2=\fivesy
    \textfont3=\tenex       \scriptfont3=\tenex
                            \scriptscriptfont3=\tenex
    \textfont\itfam=\nineit \def\it{\fam\itfam\nineit}%
    \textfont\slfam=\ninesl \def\sl{\fam\slfam\ninesl}%
    \textfont\bffam=\ninebf \scriptfont\bffam=\sixbf
                            \scriptscriptfont\bffam=\fivebf
                            \def\bf{\fam\bffam\ninebf}%
    \textfont\ssfam=\niness \scriptfont\ssfam=\sixss
                            \scriptscriptfont\ssfam=\sixss
                            \def\ss{\fam\ssfam\niness}%
    \textfont\mibfam=\ninemib \scriptfont\mibfam=\sixmib
                            \scriptscriptfont\mibfam=\sixmib
                            \def\mib{\fam\mibfam\ninemib}%
    \normalbaselineskip=12pt
    \setbox\strutbox=\hbox{\vrule height8.0pt depth3.0pt width0pt}%
    \let\big=\ninebig
    \normalbaselines\rm}

\def\eightpoint{\def\rm{\fam0\eightrm}%
    \textfont0=\eightrm      \scriptfont0=\sixrm
                             \scriptscriptfont0=\fiverm
    \textfont1=\eighti       \scriptfont1=\sixi
                             \scriptscriptfont1=\fivei
    \textfont2=\eightsy      \scriptfont2=\sixsy
                             \scriptscriptfont2=\fivesy
    \textfont3=\tenex        \scriptfont3=\tenex
                             \scriptscriptfont3=\tenex
    \textfont\itfam=\eightit \def\it{\fam\itfam\eightit}%
    \textfont\slfam=\eightsl \def\sl{\fam\slfam\eightsl}%
    \textfont\bffam=\eightbf \scriptfont\bffam=\sixbf
                             \scriptscriptfont\bffam=\fivebf
                             \def\bf{\fam\bffam\eightbf}%
    \textfont\ssfam=\eightss \scriptfont\ssfam=\sixss
                             \scriptscriptfont\ssfam=\sixss
                             \def\ss{\fam\ssfam\eightss}%
    \textfont\mibfam=\eightmib \scriptfont\mibfam=\sixmib
                             \scriptscriptfont\mibfam=\sixmib
                             \def\mib{\fam\mibfam\eightmib}%
    \normalbaselineskip=10pt
    \setbox\strutbox=\hbox{\vrule height7.0pt depth2.0pt width0pt}%
    \let\big=\eightbig
    \normalbaselines\rm}

\def\tenbig#1{{\hbox{$\left#1\vbox to8.5pt{}\right.\n@space$}}}
\def\ninebig#1{{\hbox{$\textfont0=\tenrm\textfont2=\tensy
                       \left#1\vbox to7.25pt{}\right.\n@space$}}}
\def\eightbig#1{{\hbox{$\textfont0=\ninerm\textfont2=\ninesy
                       \left#1\vbox to6.5pt{}\right.\n@space$}}}

\font\sectionfont=cmbx10
\font\subsectionfont=cmti10

\def\figurecaptionfont{\ninepoint}
\def\tablecaptionfont{\ninepoint}
\def\footnotefont{\eightpoint}


\newcount\equationno
\newcount\bibitemno
\newcount\figureno
\newcount\tableno

\equationno=0
\bibitemno=0
\figureno=0
\tableno=0


\footline={\ifnum\pageno=0{\hfil}\else
{\hss\rm\the\pageno\hss}\fi}


\def\section #1. #2 \par
{\vskip0pt plus .10\vsize\penalty-100 \vskip0pt plus-.10\vsize
\vskip 1.6 true cm plus 0.2 true cm minus 0.2 true cm
\global\def\equationlabel{#1}
\global\equationno=0
\leftline{\sectionfont #1. #2}\par
\immediate\write\terminal{Section #1. #2}
\vskip 0.7 true cm plus 0.1 true cm minus 0.1 true cm
\noindent}


\def\subsection #1 \par
{\vskip0pt plus 1.0 true cm\penalty-50 \vskip0pt plus-1.0 true cm
\vskip2.5ex plus 0.1ex minus 0.1ex
\leftline{\subsectionfont #1}\par
\immediate\write\terminal{Subsection #1}
\vskip1.0ex plus 0.1ex minus 0.1ex
\noindent}


\def\appendix #1. #2 \par
{\vskip0pt plus .10\vsize\penalty-100 \vskip0pt plus-.10\vsize
\vskip 1.6 true cm plus 0.2 true cm minus 0.2 true cm
\global\def\equationlabel{\hbox{\rm#1}}
\global\equationno=0
\leftline{\sectionfont Appendix #1. #2}\par
\immediate\write\terminal{Appendix #1. #2}
\vskip 0.7 true cm plus 0.1 true cm minus 0.1 true cm
\noindent}



\def\equation#1{$$\displaylines{\qquad #1}$$}
\def\enum{\global\advance\equationno by 1
\hfill\llap{{\rm(\equationlabel.\the\equationno)}}}
\def\noenum{\hfill}

\def\nexteq#1{\cr\noalign{\vskip#1}\qquad}


\def\ifundefined#1{\expandafter\ifx\csname#1\endcsname\relax}

\def\ref#1{\ifundefined{#1}?\immediate\write\terminal{unknown reference
on page \the\pageno}\else\csname#1\endcsname\fi}

\newwrite\terminal
\newwrite\bibitemlist

\def\bibitem#1#2\par{\global\advance\bibitemno by 1
\immediate\write\bibitemlist{\string\def
\expandafter\string\csname#1\endcsname
{\the\bibitemno}}
\item{[\the\bibitemno]}#2\par}

\def\beginbibliography{
\vskip0pt plus .15\vsize\penalty-100 \vskip0pt plus-.15\vsize
\vskip 1.2 true cm plus 0.2 true cm minus 0.2 true cm
\leftline{\sectionfont References}\par
\immediate\write\terminal{References}
\immediate\openout\bibitemlist=biblist
\frenchspacing\parindent=1.8em
\vskip 0.5 true cm plus 0.1 true cm minus 0.1 true cm}

\def\endbibliography{
\immediate\closeout\bibitemlist
\nonfrenchspacing\parindent=1.0em}


\def\figurecaption#1{\global\advance\figureno by 1
\narrower\figurecaptionfont
Fig.~\the\figureno. #1}

\def\tablecaption#1{\global\advance\tableno by 1
\vbox to 0.25 true cm { }
\centerline{\tablecaptionfont%
Table~\the\tableno. #1}
\vskip-0.4 true cm}

\def\thicktablerule{\hrule height0.8pt}
\def\thintablerule{\hrule height0.4pt}

\tenpoint

\immediate\openin\bibitemlist=biblist
\ifeof\bibitemlist\immediate\closein\bibitemlist
\else\immediate\closein\bibitemlist
\input biblist \fi


\def\thismonth{\ifcase\month\or
January\or February\or March\or April\or May\or June\or
July\or August\or September\or October\or November\or December\fi}

\input epsf
\epsfclipon



\def\rmd{{\rm d}}

\def\rme{{\rm e}}
\def\rmO{{\rm O}}



\def\proof{\noindent{\sl Proof:}\kern0.6em}

\def\frac#1#2{\hbox{$#1\over#2$}}
\def\dual{\mathstrut^*\kern-0.1em}

\def\lvec#1{\setbox0=\hbox{$#1$}
    \setbox1=\hbox{$\scriptstyle\leftarrow$}
    #1\kern-\wd0\smash{
    \raise\ht0\hbox{$\raise1pt\hbox{$\scriptstyle\leftarrow$}$}}
    \kern-\wd1\kern\wd0}
\def\rvec#1{\setbox0=\hbox{$#1$}
    \setbox1=\hbox{$\scriptstyle\rightarrow$}
    #1\kern-\wd0\smash{
    \raise\ht0\hbox{$\raise1pt\hbox{$\scriptstyle\rightarrow$}$}}
    \kern-\wd1\kern\wd0}
\def\slash#1{\setbox0=\hbox{$#1$}\setbox1=\hbox{$\kern1pt/$}
    #1\kern-\wd0\kern1pt/\kern-\wd1\kern\wd0}


\def\nab#1{{\nabla_{#1}}}
\def\nabstar#1{{\nabla\kern0.5pt\smash{\raise 4.5pt\hbox{$\ast$}}
               \kern-5.5pt_{#1}}}

\def\drvstar#1{{\partial\kern0.5pt\smash{\raise 4.5pt\hbox{$\ast$}}
               \kern-6.0pt_{#1}}}

\def\ldrvstar#1{{\lvec{\,\partial}\kern-0.5pt\smash{\raise 4.5pt\hbox{$\ast$}}
               \kern-5.0pt_{#1}}}


\def\MSbar{\overline{\rm MS\kern-0.5pt}\kern0.5pt}



\def\psibar{\overline{\psi}{\vphantom{\psi}}\kern-0.6pt}
\def\chibar{\overline{\chi}{\vphantom{\chi}}\kern-0.6pt}
\def\ren#1{#1_{\hbox{\sixrm R}}}


\def\dirac#1{\gamma_{#1}}
\def\diracstar#1#2{
    \setbox0=\hbox{$\gamma$}\setbox1=\hbox{$\gamma_{#1}$}
    \gamma_{#1}\kern-\wd1\kern\wd0
    \smash{\raise4.5pt\hbox{$\scriptstyle#2$}}}


\def\SUthree{{\rm SU(3)}}

\def\suthree{\frak{su}(3)}

\def\tr{{\rm tr}}


\def\SG{S_{\rm G}}
\def\SF{S_{\rm F}}

\def\Dw{D_{\rm w}}
\def\cG{c_{\hbox{\sixrm G}}}
\def\cF{c_{\hbox{\sixrm F}}}
\def\cX{c_{\hbox{\sixrm X}}}
\def\ps{p_{\rm s}}


\def\trans{{\Bbb T}}
\def\bstate{\Omega}
\def\states{{\cal H}}


\def\Mfld{{\cal M}}
\def\obs{{\cal O}}
\def\tauint{\tau_{\rm int}}
\def\avgavg#1{\langle\!\langle#1\rangle\!\rangle}
\def\avg#1{{\kern1.0pt\overline{\kern-1.0pt#1\kern-1.0pt}\kern1.0pt}}
\def\ssum#1{\hbox{$\sum_{#1}$}\kern1pt}
\def\SMDg{{\rm SMD}_{\gamma}}
\def\SMDs{{\rm SMD}_{0.3}}
\def\Ebar{\kern1.5pt\overline{\kern-1.5ptE\kern-0.5pt}\kern0.5pt}
\def\Qbar{\kern1.0pt\overline{\kern-1.0ptQ\kern-1.0pt}\kern1.0pt}
\def\cvec#1{\kern-0.5pt\vec{\kern0.5pt #1}}
\def\dtau{\delta\tau}
\def\Pacc{P_{\rm acc}}
\def\tacc{t_{\rm acc}}
\def\Ncnfg{N_{\rm cnfg}}
\def\kmax{k_{\rm max}}
\def\ttot{t_{\rm tot}}
\def\Gbar{\kern0.5pt\overline{\kern-0.5pt\Gamma\kern-0.5pt}\kern0.5pt}

%
\rightline{CERN-PH-TH/2011-116}
\vskip1.2cm 
\centerline{\Bigrm
Lattice QCD without topology barriers}
\vskip 0.6 true cm
\centerline{\bigrm Martin L\"uscher and Stefan Schaefer}
\vskip1.5ex
\centerline{{\it CERN, Physics Department, 1211 Geneva 23, Switzerland}}
\vskip 0.8 true cm
\thintablerule
\vskip 2.0ex
\ninepoint
\leftline{\bf Abstract}
\vskip 1.0ex\noindent
As the continuum limit is approached, lattice QCD simulations tend to
get trapped in the topological charge sectors of field space
and may consequently give biased results in practice.
We propose to bypass this problem by imposing open (Neumann) boundary
conditions on the gauge field in the time direction.
The topological charge can then flow in and out of the lattice,
while many properties of the theory (the hadron spectrum, for example) 
are not affected.
Extensive simulations of the $\SUthree$ gauge theory, 
using the HMC and the closely related SMD algorithm, 
confirm the absence of topology barriers if these boundary 
conditions are chosen.
Moreover, the calculated autocorrelation times are found to
scale approximately like the square of the inverse lattice spacing, 
thus supporting the conjecture that the HMC algorithm is in 
the universality class of the Langevin equation.
\vskip 2.0ex
\thintablerule

\tenpoint

\vskip-0.3cm

\section 1. Introduction

One of the central goals in numerical lattice QCD is the computation
of the properties of the light mesons and baryons with controlled errors. 
While the most important systematic errors in these calculations 
(finite volume and lattice spacing effects) are theoretically 
well understood, the relevant time scales in QCD simulations
remain unpredictable. 
In practice, the correctness of the simulations
within the quoted statistical errors
can therefore only be established through empirical tests
and thus only to a limited level of confidence.

In order to preserve the translation symmetry,
the lattice theory is usually set up with periodic boundary 
conditions in all space-time directions. 
A side-effect of this choice of boundary 
conditions is the emergence of disconnected topological sectors 
in the continuum limit. 
On the lattice the sectors are not strictly separated from
each other,
but the relative weight 
in the functional integral of the gauge fields ``between the sectors'' 
decreases with a high power of 
the lattice spacing [\ref{WilsonFlow}].
As a consequence,
transitions from one sector to another
tend to be suppressed in the simulations
and may eventually become so rare that a 
proper sampling of the sectors would require 
far longer runs than are practically feasible
[\ref{DelDebbioTauQ}--\ref{Villasimius}].

In this paper we address both issues, the very long
autocorrelation times caused by the emergence of the 
topological sectors and the lack of theoretical control 
over the simulations. The first of them 
we propose to avoid by imposing open boundary conditions 
on the gauge field in the time direction (see sect.~2). 
With this choice, the field space in the continuum 
theory becomes connected and the topological charge
can smoothly flow in and out of space-time through its boundaries. 
All statistically relevant parts of field space are therefore expected be
accessible to the simulation algorithms without having to cross
higher and higher topology barriers as the lattice spacing is reduced.

When properly renormalized, some algorithms may even 
converge to a well-defined stochastic process in the 
continuum limit (sect.~3). In asymptotically free theories,
such algorithms have a predictable scaling behaviour as
a function of the lattice spacing and are thus theoretically
controlled to some extent. 
The HMC algorithm [\ref{HMC}] recently turned out to be
non-renormalizable in perturbation theory
and is therefore not of this kind [\ref{NonRenHMC}],
but the algorithm may conceivably fall in the universality class of 
the Langevin equation (whose renormalizablity was
established long ago [\ref{Zinn},\ref{ZinnZwanziger}]).
The empirical tests reported in sections 4 and 5 partly serve
to verify that the topology barriers are indeed absent 
if open boundary conditions are imposed and partly to find out
whether the HMC algorithm scales like an algorithm
that integrates the Langevin equation.

\section 2. QCD with open boundary conditions

Open boundary conditions
are easily imposed in QCD and do not give rise to important
theoretical complications. While the discussion in this section
is more generally valid, 
the gauge group is taken to be $\SUthree$ from the beginning 
and we assume there is a multiplet of  
quarks in the fundamental representation of $\SUthree$.
Our notational conventions are summarized 
in appendix A.

\subsection 2.1 Boundary conditions in the continuum theory

In the continuum limit, the gauge and quark fields 
live on a four-dimensional space-time 
$\Mfld$ with Euclidean metric,
time extent $T$ and spatial size $L\times L\times L$.
Time thus runs from $0$ to $T$, while space 
is taken to be a three-dimensional torus, i.e.~all fields are required
to satisfy periodic boundary conditions in the space directions.
At time $0$ and $T$, the boundary conditions imposed
on the gauge potential $A_{\mu}(x)$ are
\equation{
   \left.F_{0k}(x)\right|_{x_0=0}=\left.F_{0k}(x)\right|_{x_0=T}=0
   \quad\hbox{for all}\quad k=1,2,3,
   \enum
}
where $F_{\mu\nu}(x)$ denotes the gauge-field tensor.
Note that these conditions preserve the gauge symmetry and
therefore do not constrain
the gauge degrees of freedom of the field.
In perturbation theory, the boundary conditions on the latter instead
derive from the gauge-fixing procedure. 
If the usual Lorentz-covariant gauge is chosen,
for example, the time and space components 
of the gauge potential are found to satisfy
Dirichlet and Neumann boundary conditions, 
respectively\kern1pt\footnote{$\dagger$}{\footnotefont%
As is the case with periodic boundary conditions,
perturbation theory in finite volume is complicated by the 
presence of non-trivial gauge-field configurations with vanishing action
(the constant Abelian fields). The remarks 
made here and below on perturbation theory
refer to the situation at $L=\infty$ and finite $T$, 
where the minimum of the action is unique up to 
gauge transformations.
}.
 
In the case of the quark and antiquark fields $\psi(x)$ and $\psibar(x)$,
we require that
\equation{
   \left.P_{+}\psi(x)\right|_{x_0=0}=\left.P_{-}\psi(x)\right|_{x_0=T}=0,
   \qquad P_{\pm}=\frac{1}{2}(1\pm\dirac{0}),
   \enum
   \nexteq{3.0ex}
   \left.\psibar(x)P_{-}\right|_{x_0=0}=\left.\psibar(x)P_{+}\right|_{x_0=T}=0.
   \enum
}
These boundary conditions are familiar from the discussion of
the QCD Schr\"odinger functional [\ref{SFunc},\ref{Sint}].
Many of the theoretical results obtained in that context can
actually be reused here. In particular, as explained in
ref.~[\ref{SFbcd}], one is practically forced to 
choose the boundary conditions (2.2),(2.3) if parity 
and the time reflection symmetry are to be preserved.
The action of the theory (without gauge fixing terms) is then given 
as usual by
\equation{
   S=-{1\over2g_0^2}\int_{\Mfld}\rmd^4x\,
   \tr\{F_{\mu\nu}(x)F_{\mu\nu}(x)\}
   +\int_{\Mfld}\rmd^4x\,
   \psibar(x)\left(\dirac{\mu}D_{\mu}+M_0\right)\psi(x),
   \enum
}
where $g_0$ and $M_0$ are the bare coupling and quark mass matrix.

\subsection 2.2 Topology of the classical field space

Since $\Mfld$ is contractible to a three-dimensional torus,
all $\SUthree$ principal bundles over $\Mfld$ are trivializable.
Smooth classical gauge potentials may therefore be assumed to 
be globally defined differentiable fields. 
In view of the non-linearity 
of the boundary conditions (2.1), the classical field space
is however not a linear space. 

We now show that the field space is connected.
Evidently, any given gauge potential $A_{\mu}(x)$ satisfying
$A_0(x)=\partial_0A_k(x)=0$ at $x_0=0$ and $x_0=T$ 
can be smoothly contracted to zero, without violating
the boundary conditions, by multiplication with a scale factor. 
These fields are therefore continuously connected to the 
classical vacuum configuration. On the other hand,
if one starts from an arbitrary field $A_{\mu}(x)$
in the classical field space,
a smooth curve of gauge transformations $\Lambda_s(x)$, $0\leq s\leq1$,
may be defined through the differential equation
\equation{
   \left(\partial_0+sA_{0}(x)\right)\Lambda_s(x)^{-1}=0,
   \qquad
   \left.\Lambda_s(x)\right|_{x_0=0}=1.
   \enum
}
When applied to the potential $A_{\mu}(x)$, the transformation
generates a curve in field space (parametrized by $s$) along which the 
field is continuously deformed to another field at $s=1$ 
with vanishing time component.
The transformed field can then be contracted to zero, as explained above,
which proves that the field space is connected.

The absence of disconnected 
topological sectors goes along with the fact that
the topological charge
\equation{
  Q=\int_{\Mfld}\rmd^4x\, q(x),
  \qquad q(x)=-{1\over32\pi^2}\,\epsilon_{\mu\nu\rho\sigma}
  \tr\{F_{\mu\nu}(x)F_{\rho\sigma}(x)\},
  \enum
}
is not quantized.
When an instanton on $\Mfld$ is contracted to the 
vacuum configuration, for example, the charge flows away through
the boundaries and $Q$ smoothly varies from $1$ to $0$.
It may be worth noting here that the massless
Dirac operator has no zero modes in the space of 
complex quark fields satisfying the boundary conditions (2.2).
Moreover, the eigenvalues $\lambda$ of the Hermitian Dirac operator
$\dirac{5}\left(\dirac{\mu}D_{\mu}+m\right)$ are all in the range
$|\lambda|>m$ (see ref.~[\ref{SFbcd}], sect.~2.2, for a proof
of these statements).
As long as the quark masses are non-negative, the quark determinant
has therefore a definite sign and never passes through zero 
even if some masses vanish.

\subsection 2.3 Renormalization and stability of the boundary conditions

The renormalization of quantum field theories on space-time 
manifolds with boundaries in general requires the usual (bulk) counterterms 
to be added to the action as well as further counterterms 
that are localized at the boundaries [\ref{Symanzik}]. 
In the present case, the symmetries of the theory, 
power counting and the fact that $\psibar\psi$ vanishes 
at the boundaries however exclude such boundary counterterms. 
The renormalization of the theory thus proceeds as in 
infinite volume by renormalizing the coupling, the quark masses 
and the fields in the correlation functions considered. 

Boundary conditions are subject to renormalization
too and sometimes require a fine-tuning of boundary counterterms.
Neumann boundary conditions in scalar field theories, 
for example, are known to be unstable under quantum fluctuations
[\ref{Symanzik}]. The situation in QCD is safe 
from this point of view, because there are no relevant or 
marginal boundary counterterms with the required symmetries.
In particular, the boundary conditions (2.1)--(2.3) are stable
under quantum fluctuations (see ref.~[\ref{SFbcd}], sect.~3,
for a broader discussion of the subject).

\subsection 2.4 Lattice formulation

The lattice theory is set up on a hypercubic lattice with spacing $a$,
time-like extent $T+a$ and spatial size $L\times L\times L$,
where $T$ and $L$ are integer multiples of $a$. Periodic
boundary conditions are imposed on all
fields in the space directions, while time runs from $0$ to 
$T$ inclusive, the terminal time-slices being the boundaries
of the lattice.

As usual the gauge and quark fields reside on the links and
points of the lattice.
In particular, the link variables $U(x,\mu)\in\SUthree$ 
live on all links $(x,\mu)$ with both endpoints in the specified
range of time. The Wilson gauge action is then given by
the sum
\equation{
   \SG={1\over g_0^2}\sum_{p}w(p)\,\tr\{1-U(p)\}
   \enum
}
over all oriented plaquettes $p$
on the lattice, $U(p)$ being the ordered product of the link variables
around $p$.
Only those plaquettes are included in the sum whose corners
are in the time interval $[0,T]$. The weight $w(p)$ is equal to $1$
except for the spatial plaquettes at time $0$ and $T$, which
have weight $\frac{1}{2}$.

In the case of the fermion fields, a possible choice of 
the action is
\equation{
   \SF=a^4\sum_{x_0=a}^{T-a}\sum_{\cvec{x}}\psibar(x)
   \left(\Dw+M_0\right)\psi(x),
   \enum
}
where
\equation{
   \Dw=
   \frac{1}{2}\dirac{\mu}\left(\nabstar{\mu}+\nab{\mu}\right)-
   \frac{1}{2}a\nabstar{\mu}\nab{\mu}
   \enum
}
denotes the Wilson--Dirac operator and the fields are assumed
to satisfy the boundary conditions (2.2),(2.3).
Since the action (2.8) depends on the quark fields at times
$0<x_0<T$ only, one may then just as well set all components
of the fields at time $0$ and $T$ to zero. The dynamical
components of the quark fields are thus those residing
in the interior of the lattice. 

The functional integral and the basic correlation functions are now
defined in the standard manner. Evidently, only the dynamical 
components of the fermion fields are integrated over
in the functional integral. 
Note that the quark determinant is a product of real factors,
one for each quark flavour, since the Wilson--Dirac operator 
is $\dirac{5}$-hermitian with the chosen boundary conditions.
The established QCD simulation algorithms can therefore be applied 
straightforwardly.

It may not be completely obvious at this point that
the fields satisfy the boundary conditions (2.1)--(2.3) in
the continuum limit.
As already mentioned in subsect.~2.3, these boundary conditions 
are stable under quantum fluctuations, i.e.~it suffices
to check that they emerge at tree-level of perturbation theory
when the lattice spacing is sent to zero. 
The explicit expression for the quark propagator 
obtained in ref.~[\ref{LWone}] and
a similar computation of the gluon propagator
in the standard covariant gauge actually show 
this to be so.

\subsection 2.5 Quantum-mechanical representation

The formulation of the lattice theory described above
admits a quantum mechanical
description in terms of a Hilbert space $\states$ of physical states
and a bounded, positive-definite transfer matrix $\trans$ [\ref{Transfer}].
In particular, the partition function of the theory is given by
the expectation value
\equation{
   {\cal Z}=\bigl(\bstate,\trans^{T/a}\bstate\bigr)
   \enum
}
of a power of the transfer matrix in a
state $\Omega\in\states$ that encodes the chosen
boundary conditions at time $0$ and $T$. 

A relatively direct way of introducing
the transfer matrix formalism starts from a representation 
of the physical states through wave functions
depending on a gauge field $V(\cvec{x},k)$ and the components
\equation{
  \chi_{-}(\cvec{x})=P_{-}\chi(\cvec{x}),\qquad
  \chibar_{+}(\cvec{x})=\chibar(\cvec{x})P_{+}
  \enum
}
of a quark field on the spatial lattice 
(see ref.~[\ref{Sint}], for example). 
The boundary state $\Omega$ is
represented by the wave function
\equation{
  \Omega(V,\chi_{-},\chibar_{+})=
  \Bigl\{\det\bigl(1+aM_0-\frac{1}{2}a^2\nabstar{k}\nab{k}\bigr)\Bigr\}^{-1}
  \enum
}
in this language, the covariant derivatives being evaluated in
presence of the gauge field $V$. Note that 
this expression is manifestly
invariant under the gauge symmetry, the lattice symmetries
and the (vector-like) flavour transformations. 
In other words, $\Omega$ has the quantum numbers of 
the vacuum state. 

Correlation functions of gauge-invariant fields 
have a quantum mechanical interpretation as well.
The two-point function of a local scalar field $\phi(x)$, for example,
is given by
\equation{
   \langle\phi(x)\phi(y)\rangle
   \mathrel{\mathop=_{x_0>y_0}}
   {1\over{\cal Z}}
   \bigl(\bstate,
   \trans^{(T-x_0)/a}\hat{\phi}(\cvec{x})
   \trans^{(x_0-y_0)/a}\hat{\phi}(\cvec{y})
   \trans^{y_0/a}\bstate\bigr),
   \enum
}
where $\hat{\phi}(\cvec{x})$ denotes the operator field
associated to $\phi(x)$ [\ref{Transfer}]. Since the transfer
matrix and the space of physical states are 
independent of the boundary conditions at time
$0$ and $T$, the hadron masses and many other physical
quantities can, in principle, be extracted from 
such correlation functions in much the same way as on lattices 
with periodic boundary conditions in time.

\subsection 2.6 On-shell O(a) improvement

The O($a$) improvement of the lattice theory follows
the lines of refs.~[\ref{SW},\ref{SFimp}]. There is
actually very little difference with respect to 
the case of the Schr\"odinger functional discussed in 
the second of these papers.
In particular, all bulk O($a$) counterterms
and their coefficients are exactly the same as those
required for the on-shell improvement of the 
theory on the infinite lattice.

We wish to emphasize at this point that a further improvement
is not needed if one is exclusively interested in the correlation
functions of fields localized far away from the boundaries
of the lattice, where the effects of the latter 
are exponentially suppressed.
The improvement of 
correlation functions involving fields 
close to or at the boundaries however requires the 
addition of the O($a$) boundary counterterms
\equation{
   \delta S_{\rm G,b}={1\over2g_0^2}(\cG-1)\sum_{\ps}
   \tr\{1-U(\ps)\},
   \enum
   \nexteq{3.0ex}
   \delta S_{\rm F,b}=(\cF-1)a^3\sum_{\cvec{x}}
   \Bigl\{\left.\psibar(x)\psi(x)\right|_{x_0=a}+
         \left.\psibar(x)\psi(x)\right|_{x_0=T-a}\Bigr\}.
   \enum
}
In these equations, $\ps$ runs over all space-like oriented plaquettes at
the boundaries of the lattice and the coefficients
\equation{
   \cX=1+\cX^{(1)}g_0^2+\cX^{(2)}g_0^4+\ldots
   \enum
}
must be adjusted so as to cancel the boundary effects of order $a$
(since the boundary conditions are not the same, there is no reason
to expect these coefficients to coincide with those needed
to improve the Schr\"odinger functional).

\subsection 2.7 Other lattice formulations of the theory

Lattice QCD with open boundary conditions can be set 
up in many different ways. Universality actually suggests
that the details of the lattice theory become
irrelevant in the continuum limit if the
gauge, space-time and flavour symmetries are respected.
Lattice formulations that
preserve chiral symmetry away from the boundaries exist as well,
but some care is required
in this case in order to guarantee the locality of the 
lattice Dirac 
operator near the boundaries [\ref{Taniguchi},\ref{SFbcd}].

\section 3. Dynamical properties of QCD simulations

The interpretation of simulation data requires
good control over the simulation dynamics.
In this section, the relevant notions are briefly discussed
and some specific issues are addressed, which arise
when studying the scaling behaviour of QCD simulations.

\subsection 3.1 Autocorrelations

QCD simulation algorithms produce random sequences of gauge-field 
configurations recursively, where the next configuration
is obtained from the current one according to some transition 
probability. The simulation algorithms considered in this paper
are the HMC algorithm [\ref{HMC}] and the closely related
SMD (stochastic molecular dynamics, or generalized HMC) 
algorithm [\ref{Horowitz}].
In both cases the simulation time is proportional
to the molecular-dynamics time in lattice units
and will, for simplicity, be identified with the latter in the following.

Let $\obs_i$ be a set of real-valued unbiased observables 
labeled by an index $i$. Their values $\obs_i(t)$ 
measured at simulation time $t$ are statistically correlated
to some extent, i.e.~the connected parts of the $n$-point 
autocorrelation functions
\equation{
   {\cal A}(t_1,\ldots,t_n)_{i_1\ldots i_n}=
   \avgavg{
   \obs_{i_1}(t_1)\ldots\obs_{i_n}(t_n)}
   \enum
}
in general do not vanish. In this equation, the bracket $\avgavg{\ldots}$
stands for the average over infinitely many statistically independent 
parallel simulations, which is the same as the average over
time translations if the simulation is ergodic.

The connected parts of the autocorrelation functions tend to fall off
exponentially at large separations in simulation time. In particular,
the two-point autocorrelation functions
\equation{
   \Gamma_{ij}(t)=\avgavg{\obs_{i}(t)\obs_{j}(0)}-
   \avgavg{\obs_i(t)}\avgavg{\obs_j(0)}
   \enum
}
can be shown to have a spectral decomposition 
of the form\kern1pt\footnote{$\dagger$}{\footnotefont%
The HMC and the SMD algorithm both evolve the gauge field 
$U(x,\mu)$ together with
its canonical momentum $\pi(x,\mu)$ (cf.~subsect.~4.1).
Equation (3.3) is partly a consequence of detailed balance 
in phase space and only holds for observables that
do not depend on the momentum.}
\equation{
   \Gamma_{ij}(t)=
   \sum_{n=0}^{\infty}
   {\rm Re}\bigl\{c_{in}c_{jn}\lambda_n^{|t|}\bigr\},
   \qquad|\lambda_n|=\rme^{-1/\tau_n},
   \enum
}
where $\tau_0\geq\tau_1\geq\ldots$ are the 
so-called exponential autocorrelation times of the
algorithm. While these are independent of the observables 
considered, the coefficients $c_{in}$ measure
how strongly the observables $\obs_i$ couple to the eigenmode
number $n$ of the transition probability.
Note that neither the spectral values $\lambda_n$ nor
the coefficients $c_{in}$ are guaranteed to be real,
except in the case of the HMC algorithm and the 
Langevin limit of the SMD algorithm.

In practice, the integrated autocorrelation times
\equation{
   \tauint(\obs_i)=\frac{1}{2}\Delta t+
   \Delta t\sum_{k=1}^{\infty}\rho_i(k\Delta t),
   \qquad
   \rho_i(t)={\Gamma_{ii}(t)\over\Gamma_{ii}(0)},
   \enum
}
of the observables of interest play an important r\^ole,
where $\Delta t$ is the separation 
in simulation time of the observable measurements.
The sum in eq.~(3.4) amounts to a numerical integration of 
the normalized autocorrelation function $\rho_i(t)$
using the trapezoidal rule. In particular,
in the Langevin limit of the SMD algorithm or if 
the HMC algorithm is used, the formula
\equation{
   \tauint(\obs_i)=
   {\sum_{n=0}^{\infty}(c_{in})^2\tau_n\over
    \sum_{n=0}^{\infty}(c_{in})^2}
   \enum
}
and thus the bound $\tauint(\obs_i)\leq\tau_0$ hold
up to integration errors.

\subsection 3.2 Topology-changing transitions  

On lattices with periodic boundary conditions,
the probability per unit simulation time for a HMC or an SMD trajectory
to pass from one topological sector to another is 
a rapidly decreasing function of the
lattice spacing [\ref{DelDebbioTauQ}--\ref{Villasimius}].
Such topology-tunneling transitions are 
non-perturbative lattice artifacts that may informally be 
described as ``an instanton falling through the lattice''.
The integrated autocorrelation time
of the topological charge consequently tends to become very large,
sometimes to the extent that the correctness of the simulation
is compromised.

With open boundary conditions, the situation is different, because
the topological charge can change smoothly along a 
molecular-dynamics trajectory by flowing in and out 
of the lattice via its boundaries.
A catastrophic slowdown of the algorithms as
in the case of periodic boundary conditions is
therefore not expected.

\subsection 3.3 Renormalizable algorithms

The $n$-point autocorrelation functions of gauge-invariant local
fields formally look like the correlation functions
in a field theory in five dimensions, where
the simulation time is the fifth space-time coordinate.
When the lattice spacing is taken to zero, the
autocorrelation functions may then conceivably have a continuum limit,
provided the fields and the parameters (of both the theory and the 
simulation algorithm) are properly renormalized. 

Algorithms that integrate the Langevin equation are known
to be renormalizable in this sense to all orders of perturbation theory [\ref{Zinn},\ref{ZinnZwanziger}]. An example of an
algorithm of this kind is provided by the $\SMDg$ algorithm
(cf.~subsect.~4.2).
The simulation time has physical dimension $[{\rm length}]^2$ in this case
and must be renormalized according to
\equation{
   t=Z_t\ren{t}/a^2
   \enum
}
where $t$ is the simulation time in lattice units, 
$Z_t(g_0)$ a renormalization constant and
$\ren{t}$ the renormalized simulation time in some physical
units. Further renormalization is not required 
apart from the usual field and parameter renormalization.

Beyond perturbation theory, the renormalizability of an algorithm
(and thus the associated scaling laws)
can break down as a result of non-perturbative lattice artifacts. 
On lattices with periodic boundary conditions,
topology-changing transitions have this effect in the case
of the $\SMDg$ algorithm. However,
if open boundary conditions are chosen, 
there is currently no reason to expect that
the renormalizability of the algorithm
does not extend to the non-perturbative level.

\subsection 3.4 Scaling behaviour of the HMC algorithm

Free-field studies of the HMC algorithm suggest
that the exponential autocorrelation times scale linearly 
(like $a^{-1}$) if the length of the molecular-dynamics 
trajectories is scaled accordingly [\ref{GHMC}].
The algorithm however
turns out to be non-renormalizable in perturbation theory 
[\ref{NonRenHMC}] and its scaling behaviour
in the presence of interactions may therefore be
completely different.

The empirical studies reported later actually show that
the HMC algorithm (on lattices with open boundary conditions)
appears to fall into the universality class of 
the Langevin equation. In particular, the autocorrelation
times scale approximately like $a^{-2}$ rather than linearly.
From this point of view, the non-renormalizability of the
HMC algorithm in perturbation theory merely reflects the fact 
that the leading-order theory is in
the wrong dynamical universality class and therefore not a suitable 
starting point for the perturbation expansion.

\subsection 3.5 Making QCD simulations safer

In practice, numerical simulations should be
much longer (by, say, a factor $100$ at least) 
than the longest exponential autocorrelation 
time $\tau_0$, as otherwise a proper sampling of
the functional integral is not guaranteed and the
simulation may consequently be biased in an unpredictable way.
Usually the integrated autocorrelation times of 
the quantities of interest are monitored,
{\it but it should be noted that
the correctness of the simulation results 
(within the estimated statistical errors)
cannot be taken for granted if 
only these autocorrelation times are
much smaller than the total simulation time.}

Integrated autocorrelation times of both physical and
other observables can in fact be very much smaller than $\tau_0$. 
In particular, the autocorrelation 
times of noisy quantities (large Wilson loops, for example)
tend to be practically unrelated to the exponential 
autocorrelation times. To illustrate this point, consider
an observable
\equation{
   \obs_0=\obs_1+c\eta,
   \enum
}
where $c$ is a constant and 
$\eta$ a statistically independent Gaussian noise with mean zero and 
unit variance. $\obs_0$ has the same expectation value as $\obs_1$
and its autocorrelation function is given by
\equation{
   \Gamma_{00}(t)=c^2\delta_{t0}+\Gamma_{11}(t).
   \enum
}
At large $c$, i.e.~when the added noise term is large,
the integrated autocorrelation time $\tauint(\obs_0)$
decreases like $1/c^2$ and can therefore be made arbitrarily 
small. Nothing is gained in this way, 
but the example shows that
integrated autocorrelation times may not be representative of
the true autocorrelations in the simulation.

\topinsert
\vbox{
\vskip0.0cm
\centerline{\epsfysize=5.0cm\epsfbox{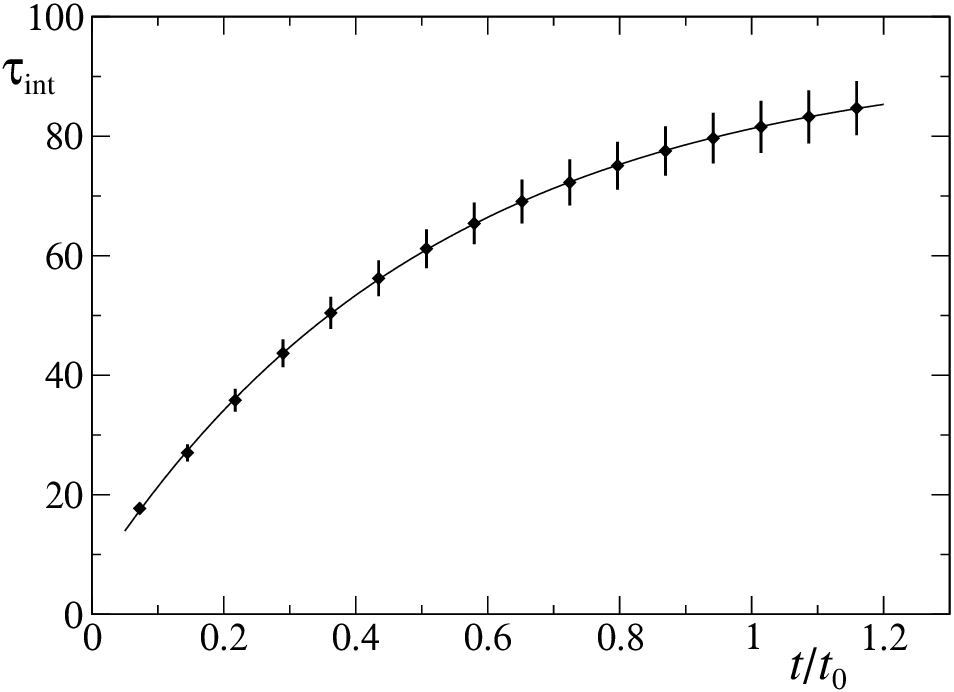}}
\vskip0.3cm
\figurecaption{%
Autocorrelation time of the density $\Ebar$ at
physical time $L/2$ in the $\SUthree$ gauge theory,
plotted as a function of the flow time $t$ (cf.~subsect.~4.3).
The simulation data (points) were obtained
on a lattice of size $32^4$ with spacing
$a=0.05$ fm and open boundary conditions, 
using the ${\rm SMD}_{0.3}$ algorithm. 
The line is a fit to the data of the form $\tauint=c_0-c_1\rme^{-c_2t}$
with $c_0\simeq94$, while the leading exponential autocorrelation
time in the even-parity sector is found to be about $100$ in 
these simulations.
}
}
\endinsert

Exponential autocorrelation times are difficult to determine
reliably if very long simulations are impractical. In this case, 
a pragmatic way to proceed is to look 
for observables with large integrated autocorrelation
times and to take the maximum of the latter as an estimate of $\tau_0$.
The observables that provide the best probes for autocorrelations
should be sensitive to the smooth modes of the gauge field, since these 
tend to be updated least efficiently.
Moreover, for the reasons given above, good probes are
likely to have small statistical fluctuations.
Quantities obtained by integrating the Wilson flow
[\ref{WilsonFlow}], such as the average action density
$\Ebar$ at positive flow time, satisfy both criteria and
are therefore recommended probes (see fig.~1).

\section 4. Numerical studies

In order to verify and complement the 
theoretical discussion in the previous sections,
we performed extensive simulations of 
the $\SUthree$ gauge theory with open boundary 
conditions. The algorithms,
observables and simulation parameters used in these studies
are described in this section.

\subsection 4.1 Simulation algorithms

Both the HMC and the SMD algorithm operate in phase space, 
i.e.~on the gauge field $U(x,\mu)$ and its canonical
$\suthree$-valued momentum field $\pi(x,\mu)$.
The $\rmO(a)$ boundary counterterm (2.14)
is not included in the Hamilton function 
\equation{
   H(\pi,U)=\frac{1}{2}(\pi,\pi)+\SG(U)
   \enum
}
of the system, partly for simplicity and partly because
the term is unimportant in the present context. 

The HMC algorithm proceeds in cycles, where in each cycle one first
chooses the momentum field randomly, with normal distribution, and then
evolves the fields according to the molecular-dynamics equations
that derive from the Hamilton function (4.1). 
In our simulations, the equations were 
integrated from molecular-dynamics time $0$ to $\tau$ 
using $n_0$ iterations of the 4th-order 
Omelyan--Mryglod--Folk (OMF) integrator 
defined through eqs.~(63) and (71) in ref.~[\ref{Omelyan}].
At the end of the evolution,
the fields are submitted to an acceptance-rejection step
that corrects for the integration errors.
This algorithm has two parameters, $\tau$ and $n_0$, and 
requires the derivative of the gauge action to be
calculated $5n_0$ times per cycle.

In the case of the SMD algorithm, one proceeds in essentially
the same way, but the momentum field is only partially refreshed
according to
\equation{
   \pi(x,\mu) \to c_1\pi(x,\mu)+c_2\upsilon(x,\mu),
   \enum
   \nexteq{2.5ex}
   c_1=\rme^{-\gamma\dtau},\qquad c_2=(1-c_1^2)^{1/2},
   \enum
}
where $\upsilon(x,\mu)$ is a randomly chosen momentum field 
with normal distribution, while $\gamma$ and $\dtau$ 
are parameters of the algorithm.
The molecular-dynamics equations are then integrated
from $0$ to $\dtau$ by applying {\it a single iteration}\/ 
of the 4th-order OMF integrator and the fields are finally 
submitted to an acceptance-rejection step.
When rejected, the fields are reset
to their values before the integration, except for a change of sign
\equation{
   \pi(x,\mu)\to-\pi(x,\mu)
   \enum
}
of the momentum field (see ref.~[\ref{JansenLiu}] for a 
straightforward proof of the correctness of the algorithm).
Note that the simulation time $t$ 
elapsed after $n$ SMD cycles 
is, by definition, equal to $n\dtau$, irrespectively of the 
rejection rate.

Since the OMF integrator is applied only once, 
the molecular-dynamics evolution time $\dtau$
is usually set to a value much smaller than $1$
in order to guarantee a high acceptance rate $\Pacc$.
Otherwise the SMD algorithm is frequently backtracking,
on average after every period of time equal to 
\equation{
   \tacc=\dtau{\Pacc\over1-\Pacc},
   \enum
}
and thus tends to become inefficient. 
With respect to the leapfrog and the 2nd-order OMF integrator,
the 4th-order OMF integrator has the advantage
that very high acceptance rates can be achieved
with a moderate computational effort.

\subsection 4.2 Stochastic equation, parameter scaling and the 
                ${\it SMD}_{\gamma}\kern-1pt$ algorithm

In the limit $\dtau\to0$, the SMD algorithm amounts to solving
the stochastic molecular-dynamics equations
\equation{
   \partial_sU_s(x,\mu)=\pi_s(x,\mu)U_s(x,\mu),
   \enum
   \nexteq{2.5ex}
   \partial_s\pi_s(x,\mu)=-T^a(\partial^a_{x,\mu}\SG)(U_s)
   -2\mu_0\pi_s(x,\mu)+\eta_s(x,\mu),
   \enum
}
where  $\eta_s$ a Gaussian random noise with mean zero and variance
\equation{
   \langle\eta_s^a(x,\mu)\eta_r^b(y,\nu)\rangle=4\mu_0\delta^{ab}\delta_{\mu\nu}
   \delta(s-r)a^{-4}\delta_{xy}.
   \enum
}
In these equations, 
the evolution time $s$ and the mass $\mu_0$ are related to the 
simulation time $t$ and the parameter $\gamma$ through
\equation{
   s=ta,\qquad \mu_0=\gamma/2a,
   \enum
}
respectively. Evidently, eqs.~(4.6),(4.7) reduce to the 
standard molecular-dynamics equations if $\mu_0$ is set to zero
(see appendix A for the definition of the derivative of the gauge
action).

When the continuum limit is approached, the scaling behaviour
of the simulation algorithms depends on how their parameters 
are scaled. The fact that
the evolution time in eqs.~(4.6),(4.7) has dimension 
[length] suggests to scale the HMC trajectory length $\tau$ 
proportionally to $1/a$ [\ref{GHMC}].
For the same reason, one can argue
that $\mu_0$ should be scaled like a physical mass
up to a logarithmically varying renormalization factor perhaps.
This choice of the parameter scaling (which, however, leads 
to non-removable ultra-violet 
singularities in perturbation theory [\ref{NonRenHMC}]) 
will be referred to as free-field scaling.

Alternatively, if $\gamma$ is held fixed,
and if $\dtau$ is such that the continuous-evolution time $\tacc$ is on 
the order of the exponential autocorrelation times (or larger),
the SMD algorithm effectively performs a numerical 
integration of the Langevin equation [\ref{NonRenHMC}]. 
For clarity, we use the acronym $\SMDg$ for the
SMD algorithm with this parameter scaling.

\subsection 4.3 Observables 

As already noted in subsect.~3.5, 
observables based on the Wilson flow 
probe the slow modes of the gauge field and
are therefore well suited for studying autocorrelations
in QCD simulations. A review of the Wilson flow
is beyond the scope of this paper and 
we merely write down the differential equation
\equation{
   \partial_tV_t(x,\mu)=
   -ag_0^2T^a(\partial^a_{x,\mu}\SG)(V_t)V_t(x,\mu),
   \qquad
   \left.V_t(x,\mu)\right|_{t=0}=U(x,\mu),
   \enum
}
that generates the flow $V_t(x,\mu)$, $t\geq0$, in the space
of gauge fields
(see refs.~[\ref{WilsonFlow},\ref{Villasimius},\ref{RenFlow}] 
for a comprehensive discussion of the flow and some of its
surprising properties).

In the course of the simulations, the observables
are evaluated at fixed separations in simulation time.
Starting from the current gauge-field configuration $U(x,\mu)$,
we first integrate the flow equation (4.10)
numerically up to some flow time $t$. The field tensor
$G_{\mu\nu}(x)$ of the gauge field $V_t(x,\mu)$ generated
in this way is defined through the clover formula, 
i.e.~through the four plaquette Wilson loops in the 
$(\mu,\nu)$-plane that start and end at $x$ 
(at the boundaries $x_0=0$ and $x_0=T$ we set $G_{0k}(x)=0$). 
The primary observables considered are then 
the time-slice averages
\equation{
   \Ebar(x_0)=-{a^3\over2L^3}\sum_{\cvec{x}}
   \tr\{G_{\mu\nu}(x)G_{\mu\nu}(x)\}
   \enum
}
of the action density and the time-slice sums
\equation{
   \Qbar(x_0)=-{a^3\over32\pi^2}\sum_{\cvec{x}}
   \epsilon_{\mu\nu\rho\sigma}
   \tr\{G_{\mu\nu}(x)G_{\rho\sigma}(x)\}
   \enum
}
of the topological charge density. Evidently, the autocorrelations
of the total charge 
\equation{
   Q=a\sum_{x_0=0}^{T}\Qbar(x_0)
   \enum
}
are studied as well. In all these equations, the 
dependence on the flow time has been suppressed 
for simplicity.

\topinsert
\newdimen\digitwidth
\setbox0=\hbox{\rm 0}
\digitwidth=\wd0
\catcode`@=\active
\def@{\kern\digitwidth}
\tablecaption{Lattice parameters} 
\vskip1.0ex
$$\vbox{\settabs\+&%
                  xxxxxx&xx&
                  xxxxxxxx&xx&
                  xxxxxxxxxxxx&xx&
                  xxxxxxxxxxxx&\cr
\thicktablerule
\vskip1.2ex
                \+& \hfill $L/a$\hfill
                 && \hfill $\beta$\hfill
                 && \hfill $a$\kern2pt[fm]\hfill
                 && \hfill $\phantom{\,^{\ast}}{t_0/a^2}\,^{\ast}$\hfill
                 &\cr
\vskip1.0ex
\thintablerule
\vskip1.2ex
  \+& \hfill $16$\hfill
  &&  \hfill $5.96$\hfill
  &&  \hfill $0.1000(6)$\hfill
  &&  \hfill $2.698(3)$\hfill
  &\cr
\vskip0.3ex
  \+& \hfill $20$\hfill
  &&  \hfill $6.09$\hfill
  &&  \hfill $0.0802(5)$\hfill
  &&  \hfill $4.203(5)$\hfill
  &\cr
\vskip0.3ex
  \+& \hfill $24$\hfill
  &&  \hfill $6.21$\hfill
  &&  \hfill $0.0667(5)$\hfill
  &&  \hfill $6.086(7)$\hfill
  &\cr
\vskip0.3ex
  \+& \hfill $32$\hfill
  &&  \hfill $6.42$\hfill
  &&  \hfill $0.0500(4)$\hfill
  &&  \hfill $11.045(15)$\hfill
  &\cr
\vskip0.3ex
  \+& \hfill $40$\hfill
  &&  \hfill $6.59$\hfill
  &&  \hfill $0.0402(3)$\hfill
  &&  \hfill $17.49(4)@@$\hfill
  &\cr
\vskip1.2ex
\thicktablerule
\vskip1.0ex
\+{\footnotefont%
$^{\ast}$ Calculated at physical time $L/2$ on the $(L/a)^4$ lattices}\cr
}
$$
\vskip-2.0ex
\endinsert

At positive flow time $t$, the expectation values of 
arbitrary (finite) products of the observables 
$\Ebar(x_0)$, $\Qbar(x_0)$ and $Q$ do not require 
renormalization and are expected to have 
a well-defined limit when the lattice spacing is taken to zero 
[\ref{WilsonFlow},\ref{RenFlow}]. 
While these expectation
values do not have any obvious interpretation in terms of glueballs
or colour flux tubes, for example, they are properties
of the continuum theory which reflect the dynamics of 
the smooth modes of the gauge field
(the smoothing radius being roughly equal to $\sqrt{8t}$). 
In particular, as explained in ref.~[\ref{WilsonFlow}],
on lattices with periodic boundary conditions, 
the topological charge $Q$ (as defined here) converges
to an integer-valued observable in the continuum limit,
which labels the topological sectors of field space.

\subsection 4.4 Lattice and simulation parameters

In table~1 we list the spatial sizes
and the inverse gauge couplings $\beta=6/g_0^2$ of the 
lattices that we have simulated. The number of lattice points
in the time direction (which is equal to $T/a+1$) coincides with $L/a$ 
in most cases, but lattices with
larger time extent have been considered too.
For the conversion to physical units, we use the 
Sommer radius $r_0=0.5$ fm [\ref{SommerScale}]
and the results obtained for $r_0/a$ by Necco and Sommer
[\ref{NeccoSommer}]. The values of the lattice spacing 
determined in this way (3rd column of table~1) are such that  
all lattices have the same physical size $L$,
as is desirable for a scaling study.

As a reference for the Wilson flow time $t$, we prefer to 
use the scale $t_0$ determined through the implicit equation
[\ref{WilsonFlow}]
\equation{
   \left\{t^2\bigl\langle\Ebar(L/2)\bigr\rangle\right\}_{t=t_0}=0.3.
   \enum
}
At flow time $t_0$, the Wilson flow has a smoothing 
range approximately equal to $r_0$, 
i.e.~this point in flow time is about where the 
non-perturbative regime sets in.
Since $L$ is quite small in physical units,
the values of $t_0/a^2$ quoted in table~1 are probably
affected by finite-volume effects and they are, in fact, 
a few percent lower
than those previously obtained in ref.~[\ref{WilsonFlow}]
at $L\simeq2.4$ fm. In the present context, the effect can
however be safely ignored since $L$ is the same on the 
lattices simulated.

\topinsert
\newdimen\digitwidth
\setbox0=\hbox{\rm 0}
\digitwidth=\wd0
\catcode`@=\active
\def@{\kern\digitwidth}
\tablecaption{Parameters of the HMC algorithm} 
\vskip-1.0ex
$$\vbox{\settabs\+&%
                  xxxxxxxxx&xx&
                  xxxxxxx&xx&
                  xxxxxxx&xx&
                  xxxxxxxx&xx&
                  xxxxxxxx&xx&
                  xxxxxxxxx&x\cr
\thicktablerule
\vskip1.2ex
                \+& \hfill Lattice\hfill
                 && \hfill $\tau$\hfill
                 && \hfill $n_0$\hfill
                 && \hfill $\Pacc$\hfill
                 && \hfill $\Delta t$\hfill
                 && \hfill $\Ncnfg$\hfill
                 &\cr
\vskip1.0ex
\thintablerule
\vskip1.2ex
  \+& \hfill $16^4$\hfill
  &&  \hfill $2.0$\hfill
  &&  \hfill $@6$\hfill
  &&  \hfill $0.953$\hfill
  &&  \hfill $@6$\hfill
  &&  \hfill $30697$\hfill
  &\cr
\vskip0.3ex
  \+& \hfill $20^4$\hfill
  &&  \hfill $2.5$\hfill
  &&  \hfill $@9$\hfill
  &&  \hfill $0.975$\hfill
  &&  \hfill $10$\hfill
  &&  \hfill $25713$\hfill
  &\cr
\vskip0.3ex
  \+& \hfill $24^4$\hfill
  &&  \hfill $3.0$\hfill
  &&  \hfill $12$\hfill
  &&  \hfill $0.979$\hfill
  &&  \hfill $15$\hfill
  &&  \hfill $25625$\hfill
  &\cr
\vskip0.3ex
  \+& \hfill $32^4$\hfill
  &&  \hfill $4.0$\hfill
  &&  \hfill $20$\hfill
  &&  \hfill $0.985$\hfill
  &&  \hfill $24$\hfill
  &&  \hfill $24041$\hfill
  &\cr
\vskip1.2ex
\thicktablerule
}
$$
\vskip-4.5ex
\endinsert

\topinsert
\newdimen\digitwidth
\setbox0=\hbox{\rm 0}
\digitwidth=\wd0
\catcode`@=\active
\def@{\kern\digitwidth}
\tablecaption{Parameters of the ${\rm SMD}_{0.3}$ algorithm} 
\vskip-1.0ex
$$\vbox{\settabs\+&%
                  xxxxxxxxxxx&xx&
                  xxxxxxxxxx&xx&
                  xxxxxxxxxx&xx&
                  xxxxxxxxx&xx&
                  xxxxxxxxxx&x\cr
\thicktablerule
\vskip1.2ex
                \+& \hfill Lattice\hfill
                 && \hfill $\dtau$\hfill
                 && \hfill $\tacc$\hfill
                 && \hfill $\Delta t$\hfill
                 && \hfill $\Ncnfg$\hfill
                 &\cr
\vskip1.0ex
\thintablerule
\vskip1.2ex
  \+& \hfill $16^4$\hfill
  &&  \hfill $0.1410$\hfill
  &&  \hfill $@516(2)$\hfill
  &&  \hfill $@5.92$\hfill
  &&  \hfill $35093$\hfill
  &\cr
\vskip0.3ex
  \+& \hfill $20^4$\hfill
  &&  \hfill $0.1128$\hfill
  &&  \hfill $@748(2)$\hfill
  &&  \hfill $@9.14$\hfill
  &&  \hfill $20209$\hfill
  &\cr
\vskip0.3ex
  \+& \hfill $24^4$\hfill
  &&  \hfill $0.0940$\hfill
  &&  \hfill $1009(3)$\hfill
  &&  \hfill $14.5@$\hfill
  &&  \hfill $20521$\hfill
  &\cr
\vskip0.3ex
  \+& \hfill $48\times24^3$\hfill
  &&  \hfill $0.0818$\hfill
  &&  \hfill $1205(2)$\hfill
  &&  \hfill $13.7@$\hfill
  &&  \hfill $70000$\hfill
  &\cr
\vskip0.3ex
  \+& \hfill $80\times24^3$\hfill
  &&  \hfill $0.0809$\hfill
  &&  \hfill $@964(2)$\hfill
  &&  \hfill $13.6@$\hfill
  &&  \hfill $40991$\hfill
  &\cr
\vskip0.3ex
  \+& \hfill $32^4$\hfill
  &&  \hfill $0.0705$\hfill
  &&  \hfill $1633(3)$\hfill
  &&  \hfill $23.7@$\hfill
  &&  \hfill $20473$\hfill
  &\cr
\vskip0.3ex
  \+& \hfill $40^4$\hfill
  &&  \hfill $0.0564$\hfill
  &&  \hfill $2368(5)$\hfill
  &&  \hfill $38.6@$\hfill
  &&  \hfill $15976$\hfill
  &\cr
\vskip1.2ex
\thicktablerule
}
$$
\vskip-3.0ex
\endinsert

The trajectory length $\tau$ in the HMC simulations
was scaled according to the free-field parameter scaling,
and the number $n_0$ of integration steps (each consisting
of one iteration of the 4th-order OMF integrator) was
then tuned to achieve fairly high acceptance rates $\Pacc$
(see table~2). In a second set of simulations, we used
the $\SMDg$ algorithm. Some experimenting suggests
that the autocorrelation times of the observables considered 
have a flat minimum near $\gamma=0.3$ and we therefore
decided to stick to this value of $\gamma$.
The other parameter of the algorithm, $\dtau$, 
was adjusted to ensure acceptance over an average simulation 
time $\tacc$ significantly
larger than the exponential autocorrelation times
(see table~3).

The observables were measured at the separations 
$\Delta t$ in simulation time quoted in tables~2 and 3.
On each lattice, a fairly large number $\Ncnfg$ of configurations were 
analyzed, the total length of the simulations 
thus being equal to $\Ncnfg\Delta t$.

\section 5. Simulation results

\vskip-4.5ex

\subsection 5.1 Scaling properties of the autocorrelation functions

While the chosen observables do not require renormalization,
the flow time at which they are evaluated should be scaled like
a physical quantity of dimension $[{\rm length}]^2$ in the 
continuum limit. In the following, 
the flow time is set to the reference time $t_0$,
the results at other values of the flow time being similar
as long as the short-time regime 
(where lattice effects are large) is avoided.

The $\SMDs$ algorithm is renormalizable to all orders
of perturbation theory since it effectively integrates the 
Langevin equation (cf.~subsects.~3.3 and 4.2).
Moreover, with open boundary conditions, the 
topology barriers that otherwise slow down the algorithm
are absent. It is therefore not unreasonable to expect that
the normalized autocorrelation functions of the 
selected observables converge to universal functions
in the continuum limit, provided the simulation time is 
scaled according to eq.~(3.6).

\topinsert
\vbox{
\vskip0.0cm
\centerline{\epsfxsize=12.0cm\epsfbox{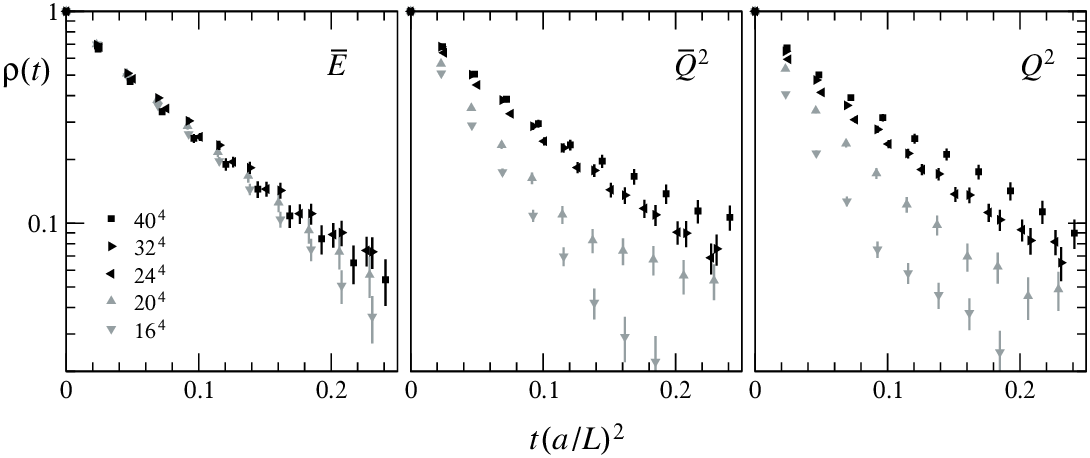}}
\vskip0.1cm
\figurecaption{%
Normalized autocorrelation functions
of the observables $\Ebar(L/2)$, $\Qbar(L/2)^2$ and $Q^2$
at flow time $t_0$, plotted as a function of the 
simulation time lag $t$ given in units of $(L/a)^2$.
The $\SMDs$ algorithm was used
all cases shown here.
For better legibility, the data points obtained on the coarsest 
lattices ($16^4$ and $20^4$) are coloured in grey, while
the black points are
those from the other lattices ($24^4$, $32^4$ and $40^4$).
}
\vskip0.0cm
}
\endinsert

The autocorrelation functions plotted in fig.~2 in fact
behave as expected if one assumes that
the renormalization constant $Z_t$ varies only slightly
on the lattices considered.
Note that all points obtained on a given lattice are statistically
correlated. In particular,
the seemingly systematic deviation of the 
measured autocorrelation functions
on the $40^4$ lattice from those on the $32^4$
and $24^4$ lattices may very well be a statistical fluctuation.
Large deviations are however seen 
in the case of the time-slice and the total topological charge
on the coarser lattices, where topology-tunneling transitions are
not totally suppressed and thus reduce the autocorrelations.
Langevin scaling then sets in as expected once these
lattice artifacts become unimportant.

On physically large lattices, the four-point autocorrelation
function of the topological charge $Q$ is dominated by its 
disconnected parts. 
The normalized two-point autocorrelation function of $Q^2$ is 
then related to the one of $Q$ through
\equation{
   \rho_{Q^2}(t)\simeq\left\{\rho_Q(t)\right\}^2.
   \enum
}
Although the simulated lattices are not very large
in physical units, we found that eq.~(5.1) is accurately satisfied.
In particular, the autocorrelation functions of $Q$ and $Q^2$ scale
in practically the same way.

\topinsert
\vbox{
\vskip0.0cm
\centerline{\epsfxsize=12.0cm\epsfbox{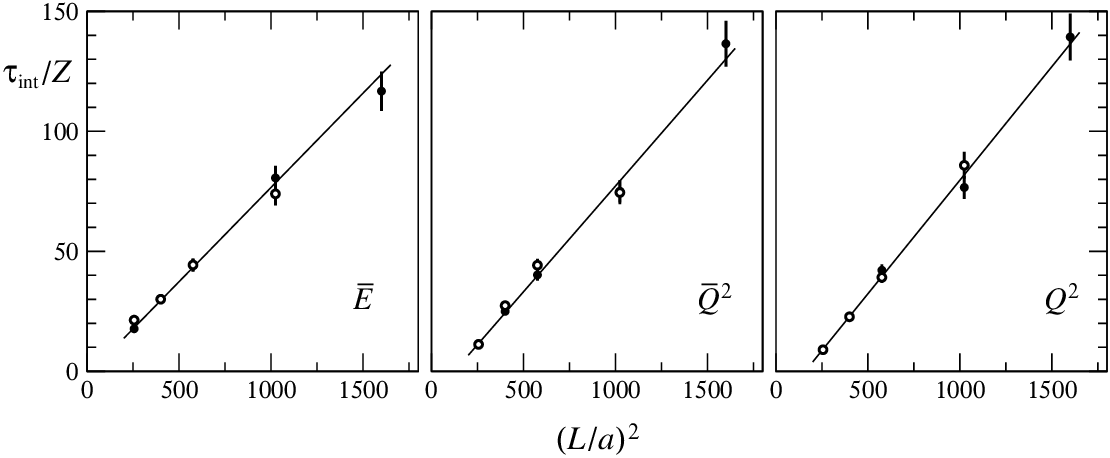}}
\vskip0.1cm
\figurecaption{%
Integrated autocorrelation times of
the observables $\Ebar(L/2)$, $\Qbar(L/2)^2$ and $Q^2$
at flow time $t_0$, as obtained
on the $(L/a)^4$ lattices using the HMC algorithm (open circles,
scale factor $Z=1.32$) and the 
$\SMDs$ algorithm (full circles, $Z=1$).
Many HMC points lie on top of the $\SMDs$ points
and thus mask the latter.
The curves are straight-line fits of the $\SMDs$ data.
}
\vskip0.0cm
}
\endinsert

\subsection 5.2 Autocorrelation times

Similarly to the energy spectrum in finite volume, the 
exponential autocorrelation times depend on the symmetry 
sector considered. In particular, eq.~(5.1) suggests 
that the longest autocorrelation time
in the odd-parity sector is larger, by a factor $2$ perhaps,
than the one in the even-parity sector.
On the basis of the data shown in fig.~2, we estimate that
the latter is about $1.2\times(r_0/a)^2$ 
(thus ranging from $30$ to $187$) in the case of the $\SMDs$ algorithm
and the $(L/a)^4$ lattices
we have simulated\kern1pt\footnote{$\dagger$}{\footnotefont%
In accordance with the conventions adopted in sect.~3,
all autocorrelation times are quoted in units of simulation
time (i.e.~molecular-dynamics 
time in lattice units).}.
As usual, such estimates should be taken with a 
grain of salt, because the slowest modes in the system
may not couple sufficiently strongly to the measured
observables for their effects to be seen in the available data.

The integrated autocorrelation times plotted in fig.~3 
and their errors were calculated 
following the lines of appendix B. 
As is evident from the figure, the autocorrelation times 
all scale linearly in $1/a^2$
and thus as expected for algorithms that integrate
the Langevin equation.
From the point of view of the continuum limit,
the intercepts at $L/a=0$ of the straight lines in fig.~3 
are $\rmO(a^2)$ lattice corrections to the Langevin scaling, while
the ratios of their slopes are universal properties of the 
simulation dynamics.

\topinsert
\vbox{
\vskip0.0cm
\centerline{\epsfxsize=12.0cm\epsfbox{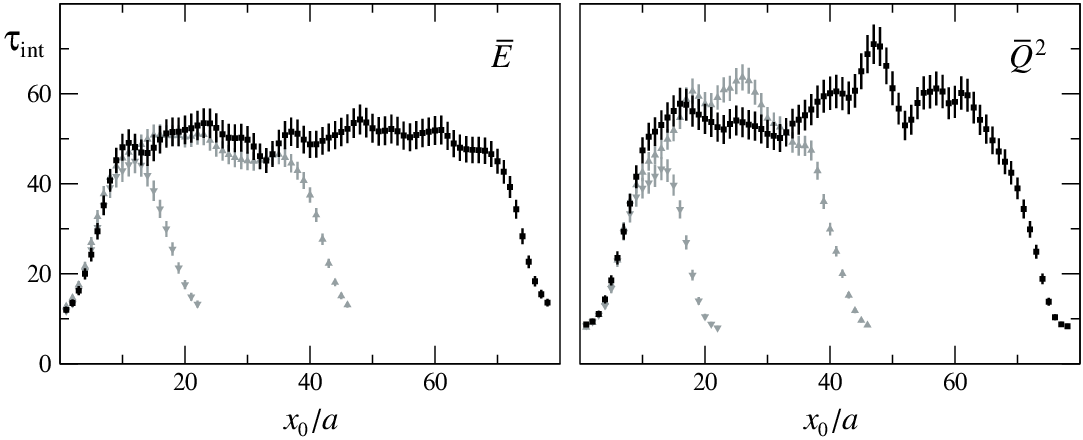}}
\vskip0.1cm
\figurecaption{%
Integrated autocorrelation times of
$\Ebar(x_0)$ and $\Qbar(x_0)^2$
at flow time $t_0$, plotted as a function of the physical time $x_0$
in lattice units.
The data were obtained on the $24^4$, $48\times24^3$ and 
$80\times24^3$ lattices using the $\SMDs$ algorithm.
For better legibility, the data points obtained on the two 
smaller lattices are coloured in grey.
}
}
\endinsert

Figure~3 also shows that the HMC algorithm (with free-field
parameter scaling) scales like the $\SMDs$ algorithm.
The matching of the autocorrelation times requires a renormalization
of the simulation time by the factor $Z\simeq1.32$, but
in terms of computer time, HMC simulations tend to be faster
than $\SMDs$ simulations, because a very accurate
integration of the molecular-dynamics equations
is not needed.

\subsection 5.3 Dependence on the time-like extent of the lattice

In practice, the time extent of the simulated lattices 
will often have to be larger
than the one of the $(L/a)^4$ lattices in our scaling studies.
Autocorrelations in general depend on the 
physical situation and thus also on the lattice geometry.
For illustration, the autocorrelation times of 
$\Ebar$ and $\Qbar^2$ calculated on three lattices with 
the same spacing and spatial size, but different time 
extent $T$, are plotted in fig.~4.
Close to boundaries of the lattice,
the autocorrelation times shown in these plots are thus
practically independent of $T$, while
well inside the lattices they rapidly converge to a 
constant value when $T$ is increased.
The behaviour of the autocorrelation times 
of these observables discussed in subsection~5.2
is therefore expected to be representative of 
the situation on larger lattices as well.

The total topological charge $Q$ is a special case,
because it can only change (at small
lattice spacings) by flowing in and out
of the lattice. In the course of a simulation, 
the measured values of $Q$ fluctuate around the origin 
with a standard deviation that increases 
proportionally to $\sqrt{TL^3}$ on large lattices.
The charge however flows through the boundaries 
with a rate proportional to $\sqrt{L^3}$ only.
The simulation time required for a significant change in $Q$
must therefore be expected to grow with $T$
(proportionally to $T$ if $Q$ performs a 
random walk). On the $24^4$, $48\times24^3$ and $80\times24^3$ lattices,
we actually find that the autocorrelation times of $Q^2$ 
($42.1(2.5)$, $113(6)$ and $148(10)$, respectively)
grow roughly linearly with $T$.

We wish to conclude this discussion by emphasizing 
that the autocorrelation times
on lattices of a given physical size are expected to scale
linearly in $1/a^2$. Independently of the chosen geometry,
the computational effort for HMC simulations
with the standard leapfrog integrator, for example, 
thus scales approximately like $1/a^7$.

\section 6. Conclusions

The theoretical and empirical results presented in this paper
show that the topology barriers in the $\SUthree$ gauge
theory can be avoided by choosing open boundary conditions
in the time direction.
Moreover, on lattices with these boundary conditions,
the HMC and the $\SMDg$ simulation algorithm both appear to fall
in the dynamical universality class of the Langevin equation,
i.e.~simulations based on these algorithms slow down proportionally
the square of the lattice spacing when the continuum limit is
approached.

In our numerical studies,
the autocorrelation times of the topological
charge (as well as those of observables unrelated to the latter) 
went up to values greater than $100$ in units of molecular-dynamics time.
While such autocorrelations may be affordable in a given case,
the experience suggests that there is ample room for 
algorithmic improvements. 
A separate treatment 
of the high-frequency and the smooth
modes of the gauge field, for example, might
be worth considering at this point.

Open boundary conditions can easily be imposed in QCD
with a non-zero number of sea quarks. We do not foresee
any technical issues when these boundary conditions are chosen,
but an interesting theoretical question is whether the 
Langevin equation remains renormalizable in the presence
of the pseudo-fermion fields that need to be introduced
to be able to simulate the theory
[\ref{UkawaFukugita},\ref{BatrouniEtAl}].

\vskip0.3ex
All simulations reported in this paper were performed on a
dedicated PC cluster at CERN. We are grateful
to the CERN management for funding this machine
and to the CERN IT Department for technical support.

\appendix A. Notational conventions

The Lie algebra $\suthree$ of $\SUthree$ 
may be identified with the linear space of all 
traceless anti-hermitian $3\times 3$ matrices.
We choose the generators $T^a$, $a=1,\ldots,8$, 
of the Lie algebra to be such that
\equation{
  \tr\{T^aT^b\}=-\frac{1}{2}\delta^{ab}.
  \enum
}
The general element $X$ of $\suthree$ is then given by
$X=X^aT^a$ with real components $X^a$
(repeated indices are automatically summed over).
The Euclidean Dirac matrices $\dirac{\mu}$, $\mu=0,\ldots,3$,
are assumed to be hermitian. 

Gauge potentials $A_{\mu}(x)$ take
values in $\suthree$ and are normalized such that the field tensor 
and the covariant derivatives that appear in the Dirac operator
are given by
\equation{
   F_{\mu\nu}=\partial_{\mu}A_{\nu}-\partial_{\nu}A_{\mu}+
   [A_{\mu},A_{\nu}],
   \enum
   \nexteq{2.5ex}
   D_{\mu}=\partial_{\mu}+A_{\mu}.
   \enum
}
On the lattice, the gauge-covariant forward and backward 
difference operators in presence of a lattice gauge field
$U(x,\mu)$ act on the quark field $\psi(x)$ according to
\equation{
  \nab{\mu}\psi(x)=
  {1\over a}\left\{U(x,\mu)\psi(x+a\hat{\mu})-\psi(x)\right\},
  \enum
  \nexteq{2.5ex}
  \nabstar{\mu}\psi(x)={1\over a}
  \left\{\psi(x)-U(x-a\hat{\mu},\mu)^{-1}\psi(x-a\hat{\mu})\right\},
  \enum
}
where $a$ denotes the lattice spacing and $\hat{\mu}$ 
the unit vector in direction $\mu$.

The scalar product of any two vector fields $\omega(x,\mu)$ and 
$\upsilon(x,\mu)$ with
values in $\suthree$ is normalized such that
\equation{
   (\omega,\upsilon)=-2a^4\sum_{x,\mu}
   \tr\{\omega(x,\mu)\upsilon(x,\mu)\}.
   \enum
}  
If ${\cal F}(U)$ is a differentiable function of the gauge field, its
derivative with respect to the link variable $U(x,\mu)$ in the 
direction of the generator $T^a$ is defined by
\equation{
   \partial^a_{x,\mu}{\cal F}(U)=
   \left.a^{-3}{\rmd\over\rmd t}{\cal F}(U_t)\right|_{t=0},
   \quad
   U_t(y,\nu)=\cases{\rme^{tT^a}U(x,\mu) & if $(y,\nu)=(x,\mu)$,\cr
                     \noalign{\vskip1.5ex}
                     U(y,\nu)            & otherwise.}
   \noenum
   \nexteq{-0.5ex}
   \enum
}
In particular, in the case of a scalar function ${\cal F}(U)$,
the combination $T^a\partial^a_{x,\mu}{\cal F}(U)$ 
is a vector field with values in $\suthree$ that transforms under
the adjoint representation of the gauge group.

\appendix B. Calculation of integrated autocorrelation times

The integrated autocorrelation times 
of the selected observables $\obs_i$ were obtained as usual
from the empirical estimates
\equation{
   \Gbar_{ii}(t)={\Delta t\over\ttot-t}\sum_{s=\Delta t}^{\ttot-t}
   \left(\obs_i(s)-\avg{\obs}_i\right)
   \left(\obs_i(s+t)-\avg{\obs}_i\right)
   \enum
}
of the autocorrelation functions $\Gamma_{ii}(t)$, where 
$\ttot=\Ncnfg\Delta t$ denotes the total simulation time
of the run and $\avg{\obs}_i$ the average of the
measured values of $\obs_i$.
In all cases, the autocorrelation functions
are found to decay exponentially at large time separations
with remarkably consistent values of the 
exponential autocorrelation times.
The estimate
\equation{
   \tauint(\obs_i)\simeq\frac{1}{2}\Delta t+
   \Delta t\sum_{k=1}^{\kmax}\bar{\rho}_i(k\Delta t),
   \qquad
   \bar{\rho}_i(t)={\Gbar_{ii}(t)\over\Gbar_{ii}(0)},
   \enum
}
therefore rapidly approaches a constant value 
when the ``summation window''
$W=\kmax\Delta t$ is sufficiently large.

On the $(L/a)^4$ lattices considered, the summation window
for even-parity observables was set to
\equation{
   W=(r_0/a)^2\times\cases{6.0 & (HMC runs),\cr
                           \noalign{\vskip1.3ex}
                           4.5 & ($\SMDs$ runs).\cr}
   \enum
} 
Given the measured exponential autocorrelation times (subsect.~5.2),
the systematic error that derives from the truncation of the sum (B.2)
is estimated to be at most $3\%$ with this choice.
The statistical errors of the autocorrelation functions and the 
integrated autocorrelation times were determined 
using the Madras--Sokal approximation [\ref{MadrasSokal}] 
(see ref.~[\ref{DDHMC}], appendix E, for a detailed description of
the procedure).

\beginbibliography


\bibitem{WilsonFlow}
M. L\"uscher,
{\it Properties and uses of the Wilson flow in lattice QCD},
JHEP 1008 (2010) 071


\bibitem{DelDebbioTauQ}
L. Del Debbio, H. Panagopoulos, E. Vicari,
{\it $\theta$-dependence of SU(N) gauge theories},
JHEP 08 (2002) 044


\bibitem{SchaeferTauQ}
S. Schaefer, R. Sommer, F. Virotta,
{\it Investigating the critical slowing down of QCD simulations},
PoS(LAT2009)032;
{\it Critical slowing down and error analysis in lattice QCD simulations},
Nucl. Phys. B845 (2011) 93


\bibitem{Villasimius}
M. L\"uscher,
{\it Topology, the Wilson flow and the HMC algorithm},\hfill\break
PoS(Lattice 2010)015


\bibitem{HMC}
S. Duane, A. D. Kennedy, B. J. Pendleton, D. Roweth,
{\it Hybrid Monte Carlo},
Phys. Lett. B195 (1987) 216.


\bibitem{NonRenHMC}
M. L\"uscher, S. Schaefer,
{\it Non-renormalizability of the HMC algorithm},
JHEP 1104 (2011) 104


\bibitem{Zinn}
J. Zinn--Justin,
{\it Renormalization and stochastic quantization},
Nucl. Phys. B275 [FS17] (1986) 135

\bibitem{ZinnZwanziger}
J. Zinn--Justin, D. Zwanziger, 
{\it Ward identities for the stochastic quantization of gauge fields},
Nucl. Phys. B295 [FS21] (1988) 297


\bibitem{SFunc}
M. L\"uscher, R. Narayanan, P. Weisz, U. Wolff,
{\it The Schr\"odinger functional:
a renormalizable probe for non-Abelian gauge theories},
Nucl. Phys. B384 (1992) 168

\bibitem{Sint}
S. Sint,
{\it On the Schr\"odinger functional in QCD},
Nucl. Phys. B421 (1994) 135

\bibitem{SFbcd}
M. L\"uscher,
{\it The Schr\"odinger functional in lattice QCD with exact chiral symmetry},
JHEP 0605 (2006) 042


\bibitem{Symanzik}
K. Symanzik,
{\it Schr\"odinger representation and Casimir effect in
renormalizable quantum field theory},
Nucl. Phys. B190 [FS3] (1981) 1


\bibitem{LWone}
M. L\"uscher, P. Weisz,
{\it O(a) improvement of the axial current in lattice QCD to
one loop order of perturbation theory},
Nucl. Phys. B479 (1996) 429


\bibitem{Transfer}
M. L\"uscher,
{\it Construction of a selfadjoint, strictly positive transfer
matrix for Euclidean lattice gauge theories},
Commun. Math. Phys. 54 (1977) 283


\bibitem{SW}
B. Sheikholeslami, R. Wohlert, 
{\it Improved continuum limit lattice action for QCD with Wilson fermions},
Nucl. Phys. B259 (1985) 572

\bibitem{SFimp}
M. L\"uscher, S. Sint, R. Sommer, P. Weisz,
{\it Chiral symmetry and O(a) improvement in lattice QCD},
Nucl. Phys. B478 (1996) 365


\bibitem{Taniguchi}
Y. Taniguchi,
{\it Schr\"odinger functional formalism with Ginsparg--Wilson fermion},
JHEP 0512 (2005) 037; 
{\it Schr\"odinger functional formalism with domain-wall fermion},
JHEP 0610 (2006) 027


\bibitem{Horowitz}
A. M. Horowitz,
{\it Stochastic quantization in phase space},
Phys. Lett. 156B (1985) 89;
{\it The second order Langevin equation and numerical simulations},
Nucl. Phys. B280 [FS18] (1987) 510;
{\it A generalized guided Monte Carlo algorithm},
Phys. Lett. B268 (1991) 247


\bibitem{GHMC}
A. D. Kennedy, B. Pendleton,
{\it Cost of the generalized Hybrid Monte Carlo algorithm for 
free field theory},
Nucl. Phys. B607 (2001) 456


\bibitem{Omelyan}
I. P. Omelyan, I. M. Mryglod, R. Folk,
{\it Symplectic analytically integrable decomposition
algorithms: classification, derivation, and application to molecular
dynamics, quantum and celestial mechanics simulations},
Comp. Phys. Commun. 151 (2003) 272


\bibitem{JansenLiu}
K. Jansen, C. Liu,
{\it Kramers equation algorithm for simulations of QCD
with two flavors of Wilson fermions and gauge group SU(2)},
Nucl. Phys. B453 (1995) 375 [E: {\it ibid.} B459 (1996) 437]


\bibitem{RenFlow}
M. L\"uscher, P. Weisz,
{\it Perturbative analysis of the gradient flow in non-Abelian gauge
theories},
JHEP 1102 (2011) 051


\bibitem{SommerScale}
R. Sommer,
{\it A new way to set the energy scale in lattice gauge theories
and its applications to the static force and $\alpha_s$ in
SU(2) Yang--Mills theory},
Nucl. Phys. B411 (1994) 839

\bibitem{NeccoSommer}
S. Necco, R. Sommer, {\it The $N_{\rm f}=0$ heavy quark potential
from short to intermediate distances},
Nucl. Phys. B622 (2002) 328


\bibitem{MadrasSokal}
N. Madras, A. D. Sokal, 
{\it The Pivot algorithm: a highly efficient Monte Carlo method 
for selfavoiding walk},
J. Stat. Phys. 50 (1988) 109

\bibitem{DDHMC}
M. L\"uscher,
{\it Schwarz-preconditioned HMC algorithm for two-flavor lattice QCD},
Comp. Phys. Commun. 165 (2005) 199


\bibitem{UkawaFukugita}
A. Ukawa, M. Fukugita,
{\it Langevin simulations including dynamical quark loops},
Phys. Rev. Lett. 55 (1985) 1854

\bibitem{BatrouniEtAl}
G. G. Batrouni, G. R. Katz, A. S. Kronfeld, G. P. Lepage,
B. Svetitsky, K. G. Wilson,
{\it Langevin simulations of lattice field theories},
Phys. Rev. D32 (1985) 2736

\endbibliography

\bye